\input epsf
\newfam\scrfam
\batchmode\font\tenscr=rsfs10 \errorstopmode
\ifx\tenscr\nullfont
        \message{rsfs script font not available. Replacing with calligraphic.}
        \def\scr{\cal}
\else   
        \font\sevenscr=rsfs7
        \font\fivescr=rsfs5
        \skewchar\tenscr='177 \skewchar\sevenscr='177 \skewchar\fivescr='177
        \textfont\scrfam=\tenscr \scriptfont\scrfam=\sevenscr
        \scriptscriptfont\scrfam=\fivescr
        \def\scr{\fam\scrfam}
        \def\cal{\scr}
\fi
\catcode`\@=11
\newfam\frakfam
\batchmode\font\tenfrak=eufm10 \errorstopmode
\ifx\tenfrak\nullfont
        \message{eufm font not available. Replacing with italic.}
        \def\frak{\it}
\else
	
	\font\sevenfrak=eufm7 \font\fivefrak=eufm5
	\textfont\frakfam=\tenfrak
	\scriptfont\frakfam=\sevenfrak \scriptscriptfont\frakfam=\fivefrak
	\def\frak{\fam\frakfam}
\fi
\catcode`\@=\active
\newfam\msbfam
\batchmode\font\twelvemsb=msbm10 scaled\magstep1 \errorstopmode
\ifx\twelvemsb\nullfont\def\Bbb{\bf}

	\message{Blackboard bold not available. Replacing with boldface.}
\else   \catcode`\@=11
        \font\tenmsb=msbm10 \font\sevenmsb=msbm7 \font\fivemsb=msbm5
        \textfont\msbfam=\tenmsb
        \scriptfont\msbfam=\sevenmsb \scriptscriptfont\msbfam=\fivemsb
        \def\Bbb{\relax\expandafter\Bbb@}
        \def\Bbb@#1{{\Bbb@@{#1}}}
        \def\Bbb@@#1{\fam\msbfam\relax#1}
        \catcode`\@=\active

\fi
        \font\eightrm=cmr8              \def\xrm{\eightrm}
        \font\eightbf=cmbx8             \def\xbf{\eightbf}
        \font\eightit=cmti10 at 8pt     \def\xit{\eightit}
        \font\eighttt=cmtt8             \def\xtt{\eighttt}
        \font\eightcp=cmcsc8
        \font\eighti=cmmi8              \def\xold{\eighti}
        \font\eightib=cmmib8             \def\xbold{\eightib}
        \font\teni=cmmi10               \def\old{\teni}
        \font\tencp=cmcsc10
        \font\tentt=cmtt10
        
        \font\twelvecp=cmcsc10 scaled\magstep1

        \font\twelvemathbf=cmmib10 at 12pt

	\font\sixteenmathbf=cmmib10 at 16pt
	
	\font\sixteenbf=cmb10 at 16pt

	 at10pt	
	\font\twelvehelvbold=phvb at12pt
	 at14pt
	 at16pt
	\font\sixteenhelvbold=phvb at16pt

\def\noblackbox{\overfullrule=0pt}
\noblackbox

\newtoks\headtext
\headline={\ifnum\pageno=1\hfill\else
	\ifodd\pageno{\eightcp\the\headtext}{ }\dotfill{ }{\old\folio}
	\else{\old\folio}{ }\dotfill{ }{\eightcp\the\headtext}\fi
	\fi}
\def\makeheadline{\vbox to 0pt{\vss\noindent\the\headline\break
\hbox to\hsize{\hfill}}
        \vskip2\baselineskip}
\newcount\infootnote
\infootnote=0
\def\foot#1#2{\infootnote=1
\footnote{${}^{#1}$}{\vtop{\baselineskip=.75\baselineskip
\advance\hsize by -\parindent\noindent{\xrm #2}}}\infootnote=0$\,$}
\newcount\refcount
\refcount=1
\newwrite\refwrite
\def\oldsize{\ifnum\infootnote=1\xold\else\old\fi}
\def\ref#1#2{
	\def#1{{{\oldsize\the\refcount}}\ifnum\the\refcount=1\immediate\openout\refwrite=\jobname.refs\fi\immediate\write\refwrite{\item{[{\xold\the\refcount}]} 
	#2\hfill\par\vskip-2pt}\xdef#1{{\noexpand\oldsize\the\refcount}}\global\advance\refcount by 1}
	}
\def\refout{\catcode`\@=11
        \xrm\immediate\closeout\refwrite
        \vskip2\baselineskip
        {\noindent\twelvecp References}\hfill\vskip\baselineskip
        \baselineskip=.75\baselineskip
        \input\jobname.refs
        \baselineskip=4\baselineskip \divide\baselineskip by 3
        \catcode`\@=\active\rm}

\def\hepth#1{\href{http://xxx.lanl.gov/abs/hep-th/#1}{{\xtt hep-th/#1}}}
\def\jhep#1#2#3#4{\href{http://jhep.sissa.it/stdsearch?paper=#2\%28#3\%29#4}{J. High Energy Phys. {\xbold #1#2} ({\xold#3}) {\xold#4}}}

\def\CQG#1#2#3{Class. Quantum Grav. {\xbold#1} ({\xold#2}) {\xold#3}}
\def\IJMPA#1#2#3{Int. J. Mod. Phys. {\xbf A}{\xbold#1} ({\xold#2}) {\xold#3}}

\def\JHEP{\jhep}

\def\NPB#1#2#3{Nucl. Phys. {\xbf B}{\xbold#1} ({\xold#2}) {\xold#3}}
\def\NPPS#1#2#3{Nucl. Phys. Proc. Suppl. {\xbold#1} ({\xold#2}) {\xold#3}}
\def\PLB#1#2#3{Phys. Lett. {\xbf B}{\xbold#1} ({\xold#2}) {\xold#3}}
\def\PRD#1#2#3{Phys. Rev. {\xbf D}{\xbold#1} ({\xold#2}) {\xold#3}}

\def\PRSA#1#2#3{Proc. Royal Soc. {\xbf A}{\xbold#1} ({\xold#2}) {\xold#3}}
\newcount\sectioncount
\sectioncount=0
\def\section#1#2{\global\eqcount=0
	\global\subsectioncount=0
        \global\advance\sectioncount by 1
	\ifnum\sectioncount>1
	        \vskip2\baselineskip
	\fi
	\noindent
        \line{\twelvecp\the\sectioncount. #2\hfill}
		\vskip.8\baselineskip\noindent
        \xdef#1{{\old\the\sectioncount}}}
\newcount\subsectioncount
\def\subsection#1#2{\global\advance\subsectioncount by 1
	\vskip.8\baselineskip\noindent
	\line{\tencp\the\sectioncount.\the\subsectioncount. #2\hfill}
	\vskip.5\baselineskip\noindent
	\xdef#1{{\old\the\sectioncount}.{\old\the\subsectioncount}}}
\newcount\appendixcount
\appendixcount=0
\def\appendix#1{\global\eqcount=0
        \global\advance\appendixcount by 1
        \vskip2\baselineskip\noindent
        \ifnum\the\appendixcount=1
        \hbox{\twelvecp Appendix A: #1\hfill}\vskip\baselineskip\noindent\fi
    \ifnum\the\appendixcount=2
        \hbox{\twelvecp Appendix B: #1\hfill}\vskip\baselineskip\noindent\fi
    \ifnum\the\appendixcount=3
        \hbox{\twelvecp Appendix C: #1\hfill}\vskip\baselineskip\noindent\fi}
\def\acknowledgements{\vskip2\baselineskip\noindent
        \underbar{\it Acknowledgements:}\ }
\newcount\eqcount
\eqcount=0
\def\Eqn#1{\global\advance\eqcount by 1
\ifnum\the\sectioncount=0
	\xdef#1{{\old\the\eqcount}}
	\eqno({\oldstyle\the\eqcount})
\else
        \ifnum\the\appendixcount=0
	        \xdef#1{{\old\the\sectioncount}.{\old\the\eqcount}}
                \eqno({\oldstyle\the\sectioncount}.{\oldstyle\the\eqcount})\fi
        \ifnum\the\appendixcount=1
	        \xdef#1{{\oldstyle A}.{\old\the\eqcount}}
                \eqno({\oldstyle A}.{\oldstyle\the\eqcount})\fi
        \ifnum\the\appendixcount=2
	        \xdef#1{{\oldstyle B}.{\old\the\eqcount}}
                \eqno({\oldstyle B}.{\oldstyle\the\eqcount})\fi
        \ifnum\the\appendixcount=3
	        \xdef#1{{\oldstyle C}.{\old\the\eqcount}}
                \eqno({\oldstyle C}.{\oldstyle\the\eqcount})\fi
\fi}
\def\eqn{\global\advance\eqcount by 1
\ifnum\the\sectioncount=0
	\eqno({\oldstyle\the\eqcount})
\else
        \ifnum\the\appendixcount=0
                \eqno({\oldstyle\the\sectioncount}.{\oldstyle\the\eqcount})\fi
        \ifnum\the\appendixcount=1
                \eqno({\oldstyle A}.{\oldstyle\the\eqcount})\fi
        \ifnum\the\appendixcount=2
                \eqno({\oldstyle B}.{\oldstyle\the\eqcount})\fi
        \ifnum\the\appendixcount=3
                \eqno({\oldstyle C}.{\oldstyle\the\eqcount})\fi
\fi}
\def\multi{\global\advance\eqcount by 1}
\def\multieq#1#2{\xdef#1{{\old\the\eqcount#2}}
        \eqno{({\oldstyle\the\eqcount#2})}}
\newtoks\url
\def\Href#1#2{\catcode`\#=12\url={#1}\catcode`\#=\active#2}
\def\href#1#2{{#2}}

\parskip=3.5pt plus .3pt minus .3pt
\baselineskip=14pt plus .1pt minus .05pt
\lineskip=.5pt plus .05pt minus .05pt
\lineskiplimit=.5pt
\abovedisplayskip=18pt plus 11pt minus 2pt
\belowdisplayskip=\abovedisplayskip
\hsize=14.1cm
\vsize=19cm
\hoffset=1.5cm
\voffset=1.8cm
\frenchspacing
\footline={}
\def\ss{\scriptstyle}

\def\*{\partial}
\def\punkt{\,\,.}
\def\komma{\,\,,}

\def\={\!=\!}
\def\small#1{{\hbox{$#1$}}}
\def\half{\small{1\over2}}
\def\fraction#1{\small{1\over#1}}
\def\fr{\fraction}
\def\Fraction#1#2{\small{#1\over#2}}
\def\Fr{\Fraction}
\def\eg{{\tenit e.g.}}
\def\Eg{{\tenit E.g.}}
\def\ie{{\tenit i.e.}}

\def\nlni{\hfill\break}

\def\a{\alpha}
\def\b{\beta}

\def\d{\delta}
\def\e{\varepsilon}
\def\g{\gamma}
\def\l{\lambda}

\def\G{\Gamma}
\def\L{\Lambda}

\def\Tr{Tr\,}
\def\tr{tr\,}

\def\Dslash{D\hskip-6.5pt/\hskip1.5pt}


\def\tJ{\tilde J}

\def\N{{\cal N}}

\def\Dhat{\hat D}

\def\h{\eta}

\def\th{\theta}
\def\r{\varrho}
\def\x{\chi}

\def\L{{\scr L}}

\def\STr{STr\,}


\headtext={M. Cederwall, B.E.W. Nilsson and D. Tsimpis: 
``D=10 super-Yang--Mills at $\ss O(\a'^2)$''}

\null\vskip-2cm
\line{
\epsfysize=1.7cm
\epsffile{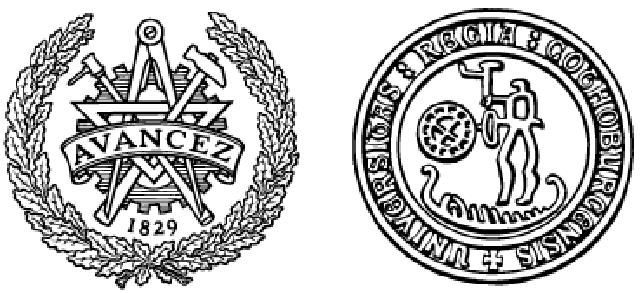}
\hfill}
\vskip-1.7cm
\line{\hfill G\"oteborg ITP preprint}
\line{\hfill\tt hep-th/0104236}
\line{\hfill April {\old2001}}
\line{\hrulefill}

\vfill

\centerline{\sixteenhelvbold D=10 super-Yang--Mills at 
{\sixteenmathbf O{\sixteenbf(}\char'13\raise2pt\hbox{$\bf '$}\raise6pt\hbox{\twelvemathbf2}{\sixteenbf)}}}  

\vskip1.6cm

\centerline{\twelvehelvbold Martin Cederwall,
	 Bengt E.W. Nilsson and Dimitrios Tsimpis}

\vskip.8cm

\centerline{\it Department of Theoretical Physics}
\centerline{\it G\"oteborg University and Chalmers University of Technology }
\centerline{\it SE-412 96 G\"oteborg, Sweden}

\vskip1.6cm

\noindent\underbar{Abstract:} Using superspace techniques, the complete
and most general action of $D=10$ super-Yang--Mills theory 
is constructed at the $\a'^2$ level. No other approximations, \eg,
keeping only a subset of the allowed derivative terms, are used.
The Lorentz structure of the $\a'^2$ corrections is
completely determined, while (depending on the gauge group) there is
some freedom in the adjoint structure, which is given by a totally 
symmetric four-index tensor. We examine the second, non-linearly
realised supersymmetry that may be present when the gauge group has
a U(1) factor, and find that the constraints from linear and
non-linear supersymmetry to a large extent coincide.
However, the additional restrictions on the adjoint structure 
of the order $\a'^2$ interactions following from the requirement of 
non-linear supersymmetry do not completely
specify the symmetrised trace prescription.
\vfill

\line{\hrulefill}
\catcode`\@=11
\line{\tentt martin.cederwall@fy.chalmers.se\hfill}
\line{\tentt bengt.nilsson@fy.chalmers.se\hfill}
\line{\tentt tsimpis@fy.chalmers.se\hfill}
\catcode`\@=\active
\eject
\ref\BI{M.~Born and L.~Infeld,
{\xit ``Foundations of the new field theory''},
\PRSA{144}{1934}{425}.}

\ref\HoweRaetzelSezginSembeddings
{P.S.~Howe, O.~Raetzel and E.~Sezgin,
{\xit ``On brane actions and superembeddings''},
\JHEP{98}{08}{1998}{011} [\hepth{9804051}].}

\ref\ACGHNNonlinear
{T.~Adawi, M.~Cederwall, M.~Holm, U.~Gran and B.E.W.~Nilsson,
{\xit ``Superembeddings, nonlinear supersymmetry and five-branes''},
\IJMPA{13}{1998}{4691} [\hepth{9711203}].}

\ref\NilssonSYM{B.E.W.~Nilsson, 
{\xit ``Off-shell fields for the 10-dimensional supersymmetric 
Yang--Mills theory''}, \xrm G\"oteborg-ITP-{\xold81}-{\xold6} ({\xold1981});
{\xit ``Pure spinors as auxiliary fields in the ten-dimensional 
supersymmetric Yang--Mills theory''},
\CQG3{1986}{{\xrm L}41}.}

\ref\GatesVashakidze{S.J.~Gates, Jr. and Sh.~Vashakidze
{\xit ``On D=10, N=1 supersymmetry, superspace geometry and 
superstring effects''},
\NPB{291}{1987}{172}.}

\ref\CNTi{M.~Cederwall, B.E.W.~Nilsson and D.~Tsimpis,
{\xit ``The structure of maximally supersymmetric Yang--Mills theory:
constraining higher-order corrections''},
\hepth{0102009}.}

\ref\Dp{M. Cederwall, A. von Gussich, B.E.W. Nilsson and A. Westerberg,
{\xit ``The Dirichlet super-three-brane in ten-dimensional type IIB 
supergravity''},
\NPB{490}{1997}{163} [\hepth{9610148}];\nlni
M. Aganagi\'c, C. Popescu and J.H. Schwarz,
{\xit ``D-brane actions with local kappa symmetry''},
\PLB{393}{1997}{311} [\hepth{9610249}];\nlni
M. Cederwall, A. von Gussich, B.E.W. Nilsson, P. Sundell
 and A. Westerberg,
{\xit ``The Dirichlet super-p-branes in ten-dimensional type IIA and IIB 
supergravity''},
\NPB{490}{1997}{179} [\hepth{9611159}];\nlni
E. Bergshoeff and P.K. Townsend, 
{\xit ``Super D-branes''},
\NPB{490}{1997}{145} [\hepth{9611173}].}

\ref\LiE{A.M. Cohen, M. van Leeuwen and B. Lisser, 
LiE v. {\xold2}.{\xold2} ({\xold1998}), 
\nlni http://wallis.univ-poitiers.fr/\~{}maavl/LiE/} 

\ref\ConvConstr{S.J.~Gates, K.S.~Stelle and P.C.~West,
{\xit ``Algebraic origins of superspace constraints in supergravity''},
\NPB{169}{1980}{347}; 
S.J. Gates and W. Siegel, 
{\xit ``Understanding constraints in superspace formulation of supergravity''},
\NPB{163}{1980}{519}.}

\ref\CNTiii{M.~Cederwall, B.E.W.~Nilsson and D.~Tsimpis,
{\xit in preparation}.}

\ref\BergshoeffFFOUR{E.~Bergshoeff, M.~Rakowski and E.~Sezgin,
{\xit ``Higher derivative super-Yang--Mills theories},
\PLB{185}{1987}{371}.}

\ref\TseytlinBIREV{A.A.~Tseytlin, 
{\xit ``Born--Infeld action, supersymmetry and string theory''},
\xrm in the Yuri Golfand memorial volume, ed. M. Shifman,
World Scientific (2000) [\hepth{9908105}]}

\ref\Kappafix{M. Aganagi\'c, C. Popescu and J.H. Schwarz,
{\xit ``Gauge-invariant and gauge fixed D-brane actions''},
\NPB{495}{1997}{145} [\hepth{9612080}].}

\ref\TseytlinSTR{A.A.~Tseytlin, 
{\xit "On the non-abelian generalization of 
Born--Infeld action in string theory"},
\NPB{501}{1997}{41} [\hepth{9701125}].}

\ref\BergshoeffKAPPA{E.A.~Bergshoeff, M.~de~Roo and A.~Sevrin,
{\xit ``On the supersymmetric non-abelian Born--Infeld action''},
\hepth{0011264}.}

\ref\DouglasNCC{M.~Douglas,
{\xit ``D-branes  and matrix theory in curved space''},
\NPPS{68}{1998}{381} [\hepth{9707228}].}

\ref\MyersDIEL{R.~Myers,
{\xit ``Dielectric branes''},
\JHEP{99}{12}{1999}{022} [\hepth{9910053}].}

\ref\WittenDBRANES{E. Witten, 
{\xit ``Bound states of strings and p-branes''},
\NPB{460}{1996}{335} [\hepth{9510135}].}

\ref\SevrinFSIX{A. Sevrin, J. Troost and W. Troost,
{\xit ``The non-abelian Born--Infeld action at order $\ss F^6$''},
\hepth{0101192}.}

\ref\TaylorNONSTR{A. Hashimoto and W. Taylor IV,
{\xit ``Fluctuation spectra of tilted and intersecting D-branes 
from the Born--Infeld action''},
\NPB{503}{1997}{193} [\hepth{9703217}].}

\ref\CGNN{M. Cederwall, U. Gran, M. Nielsen and B.E.W. Nilsson, 
{\xit ``Manifestly supersymmetric M-theory''}, 
\JHEP{00}{10}{2000}{041} [\hepth{0007035}];
{\xit ``Generalised 11-dimensional supergravity''}, \hepth{0010042}.}

\ref\SuperYM{L. Brink, J.H. Schwarz and J. Scherk, 
{\xit ``Supersymmetric Yang--Mills theories''},
\NPB{121}{1977}{77}.}

\ref\BaggerGalperin{J. Bagger and A. Galperin,
{\xit ``A new Goldstone multiplet for partially broken supersymmetry''},
\PRD{55}{1997}{1091} [\hepth{9608177}].}

\ref\HoweSezgin{P.S.~Howe and E.~Sezgin,
{\xit ``Superbranes''},
\PLB{390}{1997}{133} [\hepth{9607227}].}

\ref\NonlinearSS{P.S. Howe, O. Raetzel and E. Sezgin,
{\xit ``On brane actions and superembeddings''},
\JHEP{98}{08}{1998}{011} [\hepth{9804051}];\nlni
T. Adawi, M. Cederwall, U. Gran, M. Holm and B.E.W. Nilsson, 
{\xit ``Superembeddings, non-linear supersymmetry and 5-branes''},
\IJMPA{13}{1998}{4691} [\hepth{9711203}];\nlni 
P. Pasti, D. Sorokin and M. Tonin,
{\xit ``Superembeddings, partial supersymmetry breaking and superbranes''},
\NPB{591}{2000}{109} [\hepth{0007048}].} 

\ref\NilssonSixDSYM{B.E.W. Nilsson, 
{\xit ``Superspace action for a 6-dimensional non-extended supersymmetric
Yang--Mills theory''},
\NPB{174}{1980}{335}.}

\ref\Kitazawa{Y. Kitazawa, 
{\xit ``Effective lagrangian for open superstring from five point function''}, 
\NPB{289}{1987}{599}.} 

\ref\SevrinABI{L.~de Fosse, P.~Koerber and A.~Sevrin,
{\xit ``The uniqueness of the abelian Born--Infeld action''},
\hepth{0103015}.}

\ref\FradkinTseytlinBI{E.S.~Fradkin and A.A.~Tseytlin,
{\xit ``Non-linear electrodynamics from quantized strings''},
\PLB{163}{1985}{123}.}

\ref\TseytlinBIDerF{A.A.~Tseytlin,
{\xit ``Renormalization of Mobius infinities and partition function 
representation for string theory effective action''},
\PLB{202}{1988}{81}.}

\ref\AndreevTseytlinBIDerCorr{O.D.~Andreev and A.A.~Tseytlin,
{\xit ``Partition function representation for the open superstring 
effective action: cancellation of Mobius infinities and derivative 
corrections to Born--Infeld lagrangian''},
\NPB{311}{1988}{205}.}

\ref\CallanBIDerF{A.~Abouelsaood, C.~Callan, C.~Nappi and S.~Yost,
{\xit ``Open string in backgroud gauge fields''},
\NPB{280}{1987}{599}.}

\ref\WessBagger{J.~Wess and J.~Bagger,
{\xit ``Supersymmetry and supergravity''},
Princeton Univ. Press (1992).}

\ref\OkawaDerCorrBI{Y.~Okawa,
{\xit ``Derivative corrections to Dirac--Born--Infeld lagrangian
and non-commutative gauge theory''},
\hepth{9909132}.}

\ref\CornalbaAbelBICorr{L.~Cornalba,
{\xit ``Corrections to the abelian Born--Infeld action arising 
from noncommutative geometry''},
\JHEP{00}{09}{2000}{017} [\hepth{9912293}].}

\ref\CornalbaNonabelBI{L.~Cornalba,
{\xit ``On the general structure of the non-abelian Born--Infeld action''},
\hepth{0006018}.}

\ref\GrossWitten{D. Gross and E. Witten, 
{\xit ``Superstring modifications of Einstein's equations''},
\NPB{277}{1986}1.}

\ref\CNTHigherOrder{M. Cederwall, B.E.W. Nilsson and D. Tsimpis,
{\xit work in progress}.}

\ref\CNTCohomology{M. Cederwall, B.E.W. Nilsson and D. Tsimpis,
{\xit ``Spinorial cohomology and maximally supersymmetric theories''},
to appear on hep-th.}

\ref\SeibergWittenNonComm{N. Seiberg and E. Witten,
{\xit ``String theory and noncommutative geometry''},
\JHEP{99}{09}{1999}{032} [\hepth{9908142}].} 

\ref\WyllardDbraneCorr{N. Wyllard,
{\xit ``Derivative corrections to D-brane actions 
with constant background fields''},
\NPB{598}{2001}{247} [\hepth{0008125}].}


\section\introduction{Introduction}One of the most 
fascinating results in connection with open strings and
D-branes is the discovery [\FradkinTseytlinBI,\CallanBIDerF]
 that for an abelian and constant 
field strength the  $\a'$ expansion at string 
tree level can be summed up to yield the 
Born--Infeld theory [\BI]. The appearence of the Born--Infeld lagrangian can 
also be
understood in other ways, \eg, through the superembedding formalism 
[\HoweRaetzelSezginSembeddings] or 
as a direct consequence of requiring the existence of 
deformed BPS conditions [\SevrinABI]. This latter approach  probably 
leads to very similar if not identical results as demanding linear
supersymmetry in ten dimensions. 
However, if one relaxes the condition of constant field strength the theory
becomes substantially more complicated. A closed form for the lagrangian
is not known in this case and the action has to be obtained order by order 
in the number of derivatives correcting the original Born--Infeld action 
[\GrossWitten,\CallanBIDerF,\TseytlinBIDerF,\AndreevTseytlinBIDerCorr]. 
For a review of these 
and other results regarding the role of the Born--Infeld theory 
in string theory,
see ref. [\TseytlinBIREV]. 
Further interesting results in this direction using open-string-related 
quantities [\SeibergWittenNonComm] can be found in refs.
[\CornalbaAbelBICorr,\WyllardDbraneCorr]. 

The fact that non-abelian gauge theories
are expected to describe a stack of coincident D-branes [\WittenDBRANES] 
raises many questions concerning the
non-abelian generalisation of the abelian results alluded to above. 
As noticed by Tseytlin [\TseytlinSTR] a large class of the derivative
terms is accounted for by using a non-abelian Born--Infeld action 
defined in terms of a symmetrised trace, $STr$, over the adjoint
gauge group indices. 
In the non-abelian case, however, there are to our knowledge no results 
indicating that the action of the full theory, without any approximations,
 should be  expressible to all orders in 
 $\a'$ in closed form
similar to 
the abelian Born--Infeld theory. Therefore it is of some interest
to derive the full non-abelian D-brane action order by order in the $\a'$
expansion. Doing this directly from string theory
quickly becomes rather difficult and 
it would be useful to find other means of obtaining
these results. There are several ideas on the market, most of which 
are in one way or another
related to supersymmetry. In the abelian case, many systems with
Born--Infeld type actions and less than maximal linear supersymmetry
have also a non-linear supersymmetry that can 
be seen to follow by a Goldstone mechanism [\BaggerGalperin] from a theory
in which all supersymmetries are linearly realised. Also the 
superembedding formalism [\HoweRaetzelSezginSembeddings] is known to 
give rise to
 non-linear supersymmetries on the branes although in a much more indirect
way [\ACGHNNonlinear], but, on the other hand,
 in this approach  $\kappa$-symmetry arises very naturally 
[\HoweRaetzelSezginSembeddings]. 
For maximally supersymmetric D-branes, their $\kappa$-symmetric actions 
[\Dp] can easily be gauge-fixed and seen to reduce to systems with maximal
linear supersymmetry [\Kappafix] in less than ten dimensions.
 The Born--Infeld type actions obtained in this 
way can be immediately generalised  to the ten-dimensional abelian vector 
multiplet. In cases with maximal supersymmetry, however,  non-linear 
supersymmetries are harder to realise and are at this point
very poorly understood.

In the non-abelian situation it is known that linear 
supersymmetry fixes uniquely the action not only at lowest order but also at
order $\alpha'^2$ at least if one starts from  
$\STr F^4$ as done in ref. [\BergshoeffFFOUR]. The trace
$STr$ refers to the symmetrised trace 
in the fundamental representation introduced in [\GrossWitten]. 
Later it was proposed by Tseytlin [\TseytlinSTR] that the non-abelian
Born--Infeld action might be of the same form as the abelian one,
which is the case if a symmetric ordering prescription is imposed by means of 
the symmetric trace. Later work has indicated
that there are deviations from the $STr$ prescription 
[\TaylorNONSTR,\SevrinFSIX], but the situation
is still rather unclear.

One may also try to deduce the form of the non-abelian action by means of
non-abelian generalisations of other symmetries appearing in the abelian case,
\eg, the non-linear supersymmetry or the $\kappa$ symmetry. Trying to use
parameters valued in the adjoint of the gauge group
 seems rather involved
and would probably, if it could be realised,  have very interesting 
implications
in connection with non-commutative space-times and
matrix valued coordinates [\DouglasNCC] (see also the discussion in 
ref. [\MyersDIEL]). 
An attempt to implement such ideas
in the case of $\kappa$-symmetry is described in ref. [\BergshoeffKAPPA].
A different approach, not directly related to supersymmetry, 
to deduce the structure of non-abelian Born--Infeld theory
is used in refs. 
[\OkawaDerCorrBI,\CornalbaNonabelBI]. In these papers the authors exploit
instead the background invariance related to the Seiberg--Witten map
[\SeibergWittenNonComm].

In this paper we continue our investigations of the ten-dimensional non-abelian
supersymmetric Yang--Mills theory that we initiated in ref. [\CNTi].
Our approach relies on implementing only the linear supersymmetry 
in ten dimensions which is done
in a manifest fashion through the use of superspace 
(see, \eg, ref. [\WessBagger]).
By solving the 
Bianchi identities (BI) for the non-abelian superfield strength $F_{AB}$
\foot*{Here $\ss A,B,\ldots$ refer to the combined (vector, spinor) indices 
$\ss (a,\a), (b,\b),\ldots$, see ref. [\CNTi] for conventions.} 
imposing no constraints except for conventional ones 
[\NilssonSYM] \foot\dagger{This was also suggested in ref. 
[\GatesVashakidze].}, we managed in
ref. [\CNTi] to solve the BI's and rewrite them 
as a set of algebraic relations
between the ordinary physical fields and a set of pseudo-auxiliary fields
(here the prefix `pseudo' is used to emphasise the fact that these 
auxiliary fields 
are not sufficient to construct an action). Some of these relations involve 
derivatives on the physical fields which indicate that they
 will become field equations 
as soon as the superfield strength component of lowest dimension, $F_{\a\b}$,
 is given  explicitly
in terms of the physical fields. When we feed such an explicit form of
$F_{\a\b}$ through the solution to the
BI's the corresponding form of the field equations is obtained. 

Not knowing the exact form of $F_{\a\b}$, this program has to be
carried  out in an iterative fashion order by order in $\a'$. 
In order to find 
the most general form of the action compatible with linear supersymmetry
we thus have to construct all possible expressions at each order in the
$\a'$ expansion of $F_{\alpha \beta}$. At order 
$\a'^2$ this was done in ref. [\CNTi]  
proving that there is only one relevant expression in 
$$
F_{\a\b}=\fr{5!}\G_{\a\b}^{a_1\ldots a_5}J_{a_1\ldots a_5}\Eqn\dimoneF
$$
(the vector term is set to zero as a conventional constraint), 
namely\foot\ddagger{The overall sign is changed as compared to the expression
given in ref. [\CNTi].}
$$
J^A_{abcde}=-\half\a'^2 M^A{}_{BCD}(\l^B\G^f\G_{abcde}\G^g\l^C)F^D_{fg}
\punkt\Eqn\JExpansion
$$
Here $M^A{}_{BCD}$ is totally symmetric in all four indices and 
must be
constructed out of the invariant tensors
of the gauge group. As we will see in section {\old4}, 
this is a less restricted form of the (unique) symmetrised 
trace, $STr$, appearing in string theory [\GrossWitten] at order $\a'^2$. 
The above form of $F_{\a\b}$ has appeared
in the literature before [\GatesVashakidze,\BergshoeffFFOUR] but without 
proof of its unique status at this order. In the latter of these references
the argument was turned around in the sense 
that the component action was constructed 
first by means of Noether methods, and the result was subsequently used 
to derive the required form of $F_{\a\b}$. However, the 
symmetrised trace was used as input in the first step and the derivation
of (\JExpansion) was therefore less general than the one presented in 
ref. [\CNTi]. The resulting component action is derived in section {\old2},
and presented here with explicit adjoint group indices $A,B,C,\ldots$:
$$
\eqalign{
\L'&=\L'^{(0)}+\a'^2\L'^{(2)}\cr
&=-\fr4G^{Aij}G^A_{ij}+\half\x^A\Dslash\x^A\cr
&-6\a'^2M_{ABCD}\Bigl[\tr G^AG^BG^CG^D-\fr4(\tr G^AG^B)(\tr G^CG^D)\cr
&\phantom{-6\a'^2\Bigl\{}
-2G^{A\,i}{}_kG^{B\,jk}(\x^C\G_iD_j\x^D)			
+\fr2G^{A\,il}D_lG^{B\,jk}(\x^C\G_{ijk}\x^D)	\cr	
&\phantom{-6\a'^2\Bigl\{}
+\fr{180}(\x^A\G^{ijk}\x^B)(D_l\x^C\G_{ijk}D^l\x^D)		
+\Fr3{10}(\x^A\G^{ijk}\x^B)(D_i\x^C\G_jD_k\x^D)\cr		
&\phantom{-6\a'^2\Bigl\{}
+\Fr7{60}f^D{}_{EF}G^{A\,ij}(\x^B\G_{ijk}\x^C)(\x^E\G^k\x^F)\cr		
&\phantom{-6\a'^2\Bigl\{}
-\fr{360}f^D{}_{EF}G^{A\,ij}(\x^B\G^{klm}\x^C)(\x^E\G_{ijklm}\x^F)\Bigr]
	+O(\a'^3)\punkt\cr
}\Eqn\FinalLagrangianPreview
$$
The trace $tr$ refers to the structure of the Lorentz indices. Note also that
it is only in the last two terms that the non-abelian structure is crucial
since $M^A{}_{BCD}$ is non-zero also in the abelian case. As discussed in
section {\old5}, in the non-abelian case
the number of independent structures in $M^A{}_{BCD}$ 
depends on the gauge group.

This paper is organised as follows. In section {\old2} we insert 
eq. (\JExpansion)
into the equations obtained in ref. [\CNTi] from the BI's. The result is the
equations of motion, which in section {\old3} are integrated to
the complete action at order $\a'^2$ including
all fermionic terms required by
supersymmetry at this order. The answer agrees exactly with previously
derived terms in, \eg, ref. [\BergshoeffFFOUR] 
where however the quartic fermion terms were
not given. In section {\old4} we present the form of the supersymmetry 
transformations to the order relevant for our purposes. We also 
show that there exists an $\a'^2$-corrected non-linear supersymmetry 
provided
the gauge group has a U(1) factor and the parameter takes values only in this 
factor. The corrections at the level
considered are exactly the ones responsible for turning this symmetry
from an abelian one into an ordinary second supersymmetry.
Interestingly enough, we find that also
the fields in the semi-simple part of the gauge group transform under this
non-linear supersymmetry. However, since the parameter is abelian this
symmetry does not have any implications for  questions
concerning non-commutative space-times. Section {\old5} 
is devoted to an analysis
of the possible ways of constructing ${M^A}_{BCD}$, 
which turn out to be rather mildly affected by requiring the existence of
a non-linear supersymmetry. We collect our results and some additional 
comments in section {\old6}.

\section\Interactions{Interactions at $O(\a'^2)$ from superspace}In order to 
derive the field equations for $A_a$ and $\lambda^{\a}$ we must 
insert (suppressing both $\a'$ and ${M^A}_{BCD}$ in the following)
$$
J_{abcde}=10(\l\G_{[abc}\l)F_{de]}
+\half(\l\G_{abcde}{}^{fg}\l)F_{fg}
=-\half(\l\G^f\G_{abcde}\G^g\l)F_{fg}\Eqn\Jsecondorder
$$
into the following exact expressions  
obtained in ref. [\CNTi] 
by solving the superspace BI's:
$$
\eqalign{
0&=D^bF_{ab}-\l\G_a\l-8D^bK_{ab}+36w_a-\Fr43\{\l,\tJ_a\}
-2\tJ_b\G_a\tJ^b+\fr{140\cdot3!}\tJ_{bcd}\G_a\tJ^{bcd}\cr
&\qquad+\fr{42}[K_{bcde},J_a{}^{bcde}]
	+\fr{42\cdot4!}[D^fJ_{fbcde},J_a{}^{bcde}]
}\Eqn\AEOM
$$
and
$$
0=\Dslash\l-30\psi+\Fr43D^a\tJ_a
	+\Fr5{126\cdot5!}	
	\G^{abcde}[\l,J_{abcde}]\punkt\Eqn\LambdaEOM
$$ 
Apart from $F_{ab}$ and $\l^{\a}$, 
the quantities appearing in these equations all arise, as 
explained in ref. [\CNTi], in the $\theta$ expansion of 
$J_{abcde}$; $\tilde{J}'s$ at first,
$K's$  at second, $\psi$ at third, and $\omega$ at fourth order in $\theta$. 
Explicitly,
their precise relations to $J_{abcde}$ are given by
$$
\eqalign{
\tJ_a&=\fr{1680}\G^{bcde}DJ_{bcdea}\komma\cr
\tJ_{abc}&=-\fr{12}\G^{de}DJ_{deabc}-\fr{224}\G_{[ab}\G^{defg}DJ_{|defg|c]}
	\komma\cr
\tJ_{abcde}&=DJ_{abcde}+\Fr56\G_{[ab}\G^{fg}DJ_{|fg|cde]}
	+\fr{24}\G_{[abcd}\G^{fghi}DJ_{|fghi|e]}\komma\cr
}\Eqn\TildeJs
$$
 $$
\eqalign{
K_{ab}&=\fr{5376}(D\G^{cde}D)J_{cdeab}\komma\cr
K_{abcd}&=\fr{480}(D\G_{[a}{}^{fg}D)J_{|fg|bcd]}\komma\cr
}\Eqn\KintermsofJ
$$
$$
\psi_\a=-\fr{840\cdot3!\cdot5!}		
	\G_{abc}{}^{\b\g}\G_{de\,\a}{}^\d D_{[\b}D_\g D_{\d]}
	J^{abcde}\komma\eqn
$$
and finally
$$
w_a=\fr{4032\cdot4!\cdot5!}\G^{[\a\b}_{abc}\G^{\g\d]}_{def}
	D_\a D_\b D_\g D_\d J^{bcdef}\punkt\eqn
$$

Here we have adopted the notation that keeps track of $M^A{}_{BCD}$ 
and possible 
structure constants ${f^A}_{BC}$ 
without having to write them out explicitly:
$\l\G^{(n)}\l$ means $\fr{2}{f^A}_{BC}\l^B\G^{(n)}\l^C$ 
and makes sense only for 
$n=1$ and $n=5$, while $\{\l,\tJ_a\}$ instead means 
${f^A}_{BC}\l^B\tJ^C_a$. In the case of two bosonic quantities, 
or one fermionic and one bosonic quantity, the 
anti-commutator indicated by the curly bracket is exchanged for an ordinary
commutator bracket as in \eg\ $\G^{abcde}[\l,J_{abcde}]$.
At some places in the formul\ae\ below we use the curly
bracket notation just as a way to keep track of structure constants.
\Eg, the very last term
in $K_{ab}$ given below in eq. ({\old2}.{\old11}) should be read as 
$-\Fr3{448}{M^A}_{BCD}f^D{}_{EF}(\l^B\G_{ab}{}^i\l^C)(\l^E\G_i\l^F)$. 

Before we start analysing the consequences of eq. (\Jsecondorder),
we must make sure that at order $\theta$ in $J_{abcde}$, 
$$
DJ_{abcde}=\tJ_{abcde}+10\G_{[ab}\tJ_{cde]}+5\G_{[abcd}\tJ_{e]}\komma\eqn
$$
the field
in the irreducible representation (00030) vanishes [\CNTi]: \ie, 
that $\tJ_{abcde}=0$, where $\tJ_{abcde}$ is $\G$-traceless,
must follow as a direct consequence of (\Jsecondorder).
Computing the result of acting with
a fermionic covariant derivative on (\Jsecondorder) and 
using the lowest order relations
$D_\a\l^\b=\half(\G^{ij}){}_\a{}^\b F_{ij}$ and 
$D_\a F_{ab}=2(\G_{[a}D_{b]}\l)_\a$, 
we find trivially that the representation (00030) does not occur
in the tensor product of the constituent fields, which implies that 
the condition is satisfied modulo terms of order
$\a'^4$ and higher. 

We can now proceed  to derive the lagrangian 
at order $\a'^2$.
The calculation can be simplified as follows. We start from the 
$\l$ equation above, obtain its explicit form to
order $\a'^2$, and derive the action from which it follows. The pure $F$
terms not obtainable this way can be derived by 
acting with a fermionic derivative on the  field equation for
$\l^{\a}$ keeping only the pure $F^4$ terms.
Thus we will actually not make use of the $A_a$ equation (\AEOM)
at any point in this paper. 

Concentrating on eq. (\LambdaEOM) we first rewrite it using the
following equation also obtained in ref. [\CNTi]:
$$
\G^{ab}DK_{ab}=-\Fr{225}2\psi+\Fr52D^a\tJ_a
	+\fr{2016}\G^{abcde}[\l,J_{abcde}]\punkt\eqn
$$
Elimination of $\psi$ gives
$$
0=\Dslash\l+\Fr23D^a\tJ_a+\Fr4{15}\G^{ab}DK_{ab}
	+\fr{42\cdot5!}	
	\G^{abcde}[\l,J_{abcde}]\punkt\Eqn\LambdaEOM
$$

When acting with  spinor 
derivatives on $J_{abcde}$ to derive the expressions for
$\tJ_a$ and $K_{ab}$, we find that terms of different 
powers of $\a'$ are produced.
Higher order terms arise, \eg, 
by using equations like 
$\Lambda_{ab}=F_{ab}-{28\over5}K_{ab}$, where $\Lambda_{ab}$ comes from
expanding $D_{\a}\l^{\b}$ in terms of tensors,
$D_\a\l^\b=\Lambda\d_\a{}^\b+\half(\G^{ij}){}_\a{}^\b\Lambda_{ij}
+\fr{24}(\G^{ijkl}){}_\a{}^\b\Lambda_{ijkl}$,
and the leading term in $K_{ab}$ is of two higher powers of $\a'$ 
than $F_{ab}$.
Dropping all but the leading terms in $\a'$ we find:
$$
\eqalign{
\tJ_a&=-\Fr3{70}F^{ij}F^{kl}\G_{aijkl}\l
	+\Fr9{35}F_a{}^iF^{jk}\G_{ijk}\l\cr
	&-\fr5F_{ij}F^{ij}\G_a\l
	+2F_{aj}F^{ij}\G_i\l\cr
	&+\fr{42}(\l\G^{ijk}\l)\G_{ijk}D_a\l
	-\fr{35}(\l\G^{ijk}\l)\G_{aij}D_k\l
	+\Fr3{35}(\l\G_a{}^{ij}\l)\G_iD_j\l\komma\cr
K_{ab}&=\Fr47F_{ij}F^{ij}F_{ab}
	-\Fr{22}7F_a{}^iF_b{}^jF_{ij}\cr
	&+\Fr{13}{28}F^{ij}(\l\G_{abi}D_j\l)
	+\Fr{25}{56}F_{ij}(\l\G_a{}^{ij}D_b\l)\cr
	&+\Fr{12}7F_a{}^i(\l\G_bD_i\l)
	+\Fr{43}{28}F_a{}^i(\l\G_iD_b\l)
	+\Fr3{28}D_aF^{ij}(\l\G_{bij}\l)\cr
	&+\Fr{11}{2688}(\l\G^{ijk}\l)\{\l,\G_{abijk}\l\}
	-\Fr3{448}(\l\G_{ab}{}^i\l)\{\l,\G_i\l\}\punkt
}\Eqn\Kab
$$

In deriving the final form of the equations of motion one needs to rearrange
a number of trilinear $\l$ terms in order to have a minimal set of
linearly independent terms. We refer to the appendix for the
Fierz identities needed.
For the complete equation of motion for $\l$ at order $\a'^2$ one finds
$$
\eqalign{
0=\Dslash\l&+\Fr{28}5F^{ij}F^{kl}\G_{ijk}D_l\l
	+24F^i{}_kF^{jk}\G_iD_j\l		\cr
&-\Fr{16}5F^{il}D_lF^{jk}\G_{ijk}\l
	+\Fr45F^{jk}D_iF_{jk}\G^i\l		\cr
&-\Fr{17}{60}(D^l\l\G_{ijk}D_l\l)\G^{ijk}\l
	-\Fr95(D_i\l\G_jD_k\l)\G^{ijk}\l	\cr
&	+\fr5(\l\G^{ijk}\l)[F_i{}^l,\G_{jkl}\l]
	+\half(\l\G^{ijk}\l)[F_{ij},\G_k\l]	\cr
&+\Fr5{48}F_{ij}\{\l,\G^{ijklm}\l\}\G_{klm}\l
	-\Fr{19}{40}F^{ij}\{\l,\G^k\l\}\G_{ijk}\l
	+F^{ij}\{\l,\G_i\l\}\G_j\l	\punkt\cr
}\Eqn\LambdaEOM
$$
Since this equation is still written in terms of superfields we need to set
$\theta=0$ in order to extract the corresponding component field equation
(which is identical to the above equation).
Having done so, we now turn to the construction of the action that
gives rise to this field equation. 

\section\TheAction{The action}To systematise 
the integration of the equation of motion to a lagrangian,
we enumerate the possible terms in the equation of motion,
$$\
\eqalign{
E_1&=F^{ij}F^{kl}\G_{ijk}D_l\l			\komma\cr
E_2&=F^i{}_kF^{jk}\G_iD_j\l			\komma\cr
E_3&=F^{il}D_lF^{jk}\G_{ijk}\l			\komma\cr
E_4&=F^{jk}D_iF_{jk}\G^i\l			\komma\cr
E_5&=(D^l\l\G_{ijk}D_l\l)\G^{ijk}\l		\komma\cr
E_6&=(D_i\l\G_jD_k\l)\G^{ijk}\l			\komma\cr
E_7&=(\l\G^{ijk}\l)[F^{lm},\G_{ijklm}\l]	\komma\cr
E_8&=(\l\G^{ijk}\l)[F_i{}^l,\G_{jkl}\l]		\komma\cr
E_9&=(\l\G^{ijk}\l)[F_{ij},\G_k\l]		\komma\cr
E_{10}&=F_{ij}\{\l,\G^{ijklm}\l\}\G_{klm}\l	\komma\cr
E_{11}&=F^{ij}\{\l,\G^k\l\}\G_{ijk}\l		\komma\cr
E_{12}&=F^{ij}\{\l,\G_i\l\}\G_j\l		\komma\cr
}\eqn
$$
and write the equation as $0=\Dslash\l+\sum_{n=1}^{12}x_nE_n$, 
with the numbers $x_n$ given in eq. (\LambdaEOM),
$(x_1,\ldots,x_{12})=(\Fr{28}5,24,-\Fr{16}5,\Fr45,-\Fr{17}{60},-\Fr95,
	0,\fr5,\fr2,\Fr5{48},-\Fr{19}{40},1)$.
We also write the lagrangian as
$$
\L=\hbox{pure $F$-terms}+\half\l\Dslash\l+\sum_{m=1}^{10}a_m\L_m\komma\eqn
$$
where
$$
\eqalign{
\L_1&=F^{ij}F^{kl}(\l\G_{ijklm}D^m\l)		\komma\cr
\L_2&=F^i{}_kF^{jk}(\l\G_iD_j\l)		\komma\cr
\L_3&=F^{il}D_lF^{jk}(\l\G_{ijk}\l)		\komma\cr
\L_4&=F^{ij}F_{ij}(\l\Dslash\l)			\komma\cr
\L_5&=F^{ij}D_lF^{kl}(\l\G_{ijk}\l)		\komma\cr
\L_6&=(\l\G^{ijk}\l)(D_l\l\G_{ijk}D^l\l)	\komma\cr
\L_7&=(\l\G^{ijk}\l)(D_i\l\G_jD_k\l)		\komma\cr
\L_8&=(\l\G^{ijk}\l)(D^l\l\G_{lijkm}D^m\l)	\komma\cr
\L_9&=F^{ij}(\l\G_{ijk}\l)\{\l,\G^k\l\}		\komma\cr
\L_{10}&=F^{ij}(\l\G^{klm}\l)\{\l,\G_{ijklm}\l\}\punkt\cr
}\eqn
$$
The system of equations for determining the $a_m$'s from the $x_n$'s
is over-determined, unless two relations hold among the $x_n$'s,
namely, $12x_5-x_6+8x_8=0$, $x_2-192x_{10}-4x_{12}=0$. This consistency
check is satisfied by the $x_n$'s in eq. (\LambdaEOM). The solution is
$$
(a_1,\ldots,a_{10})=(-\Fr7{10},12,-\Fr85,-\Fr{13}5,-\Fr{11}{10},
	-\fr{30},-\Fr95,0,\Fr{11}{20},\fr{60})\punkt\eqn
$$
The only terms in the lagrangian that are not derivable from
the equation of motion for $\l$ are the ones containing only $F$.
They are quite easily calculated by taking a spinor derivative on
the equation of motion for $\l$, keeping only pure $F$-terms,
and found to be $-6[\tr F^4-\fr4(\tr F^2)^2]$, as expected from string
theory scattering amplitudes [\GrossWitten] (the traces
are over Lorentz indices with $F$ seen as a matrix; the adjoint indices
are as usual suppressed).
The complete lagrangian at $O(\a'^2)$ is thus
$$
\eqalign{
\L=&-\fr4F^{ij}F_{ij}+\half\l\Dslash\l\cr
&+\a'^2\Bigl[-6\bigl(\tr F^4-\fr4(\tr F^2)^2\bigr)\cr
&\phantom{+\a'^2\Bigl\{}
-\Fr7{10}F^{ij}F^{kl}(\l\G_{ijklm}D^m\l)		
+12F^i{}_kF^{jk}(\l\G_iD_j\l)\cr			
&\phantom{+\a'^2\Bigl\{}
-\Fr85F^{il}D_lF^{jk}(\l\G_{ijk}\l)		
-\Fr{13}5F^{ij}F_{ij}(\l\Dslash\l)			
-\Fr{11}{10}F^{ij}D_lF^{kl}(\l\G_{ijk}\l)\cr		
&\phantom{+\a'^2\Bigl\{}
-\fr{30}(\l\G^{ijk}\l)(D_l\l\G_{ijk}D^l\l)		
-\Fr95(\l\G^{ijk}\l)(D_i\l\G_jD_k\l)\cr		
&\phantom{+\a'^2\Bigl\{}
+\Fr{11}{20}F^{ij}(\l\G_{ijk}\l)\{\l,\G^k\l\}		
+\fr{60}F^{ij}(\l\G^{klm}\l)\{\l,\G_{ijklm}\l\}\Bigr]+O(\a'^3)\punkt\cr
}\eqn
$$
Some of the terms, namely those that contain lowest order field equations,
may be removed by field redefinitions.
One may consider redefinitions of the forms
$$
\eqalign{
\l&=\x+\a'^2\bigl[\a G^{ij}G_{ij}\x+\b G^{ij}G^{kl}\G_{ijkl}\x\bigr]
\komma\cr
A_a&=B_a+\g\a'^2G^{ij}(\x\G_{aij}\x)\komma\cr
}\eqn
$$
where $G$ is the field strength of $B$.
The effect of these redefinitions of the physical component fields 
could also be obtained by redoing the  superspace calculation 
using a new 
conventional constraint in
$\G_a^{\a\b}F_{\a\b}$ corresponding to
changing the vector potential, together with a redefinition of the spinor 
that goes into $\G^{i\,\a\b}F_{i\b}$.
If one wants to continue the calculations to higher order in $\a'$, it may
be more convenient not to perform them.
The redefinitions shift the coefficients $a_m$ as
$\d a_1=\b$, $\d a_3=-2\b$, $\d a_4=\a$, $\d a_5=-2\b-\g$, $\d a_9=\half\g$.
By choosing $\a=\Fr{13}5$, $\b=\Fr7{10}$, $\g=-\Fr52$, we remove the
terms $\L_1$, $\L_4$ and $\L_5$, which are the ones containing the lowest
order equations of motion, from the lagrangian.
We then obtain the lagrangian in its simplest possible form at this level:
$$
\eqalign{
\L'&=\L'^{(0)}+\a'^2\L'^{(2)}\cr
&=-\fr4G^{ij}G_{ij}+\half\x\Dslash\x\cr
&-6\a'^2\Bigl[\tr G^4-\fr4(\tr G^2)^2\cr
&\phantom{-6\a'^2\Bigl\{}
-2G^i{}_kG^{jk}(\x\G_iD_j\x)			
+\fr2G^{il}D_lG^{jk}(\x\G_{ijk}\x)	\cr	
&\phantom{-6\a'^2\Bigl\{}
+\fr{180}(\x\G^{ijk}\x)(D_l\x\G_{ijk}D^l\x)		
+\Fr3{10}(\x\G^{ijk}\x)(D_i\x\G_jD_k\x)\cr		
&\phantom{-6\a'^2\Bigl\{}
+\Fr7{60}G^{ij}(\x\G_{ijk}\x)\{\x,\G^k\x\}		
-\fr{360}G^{ij}(\x\G^{klm}\x)\{\x,\G_{ijklm}\x\}\Bigr]+O(\a'^3)\punkt\cr
}\Eqn\FinalLagrangian
$$
The terms up to quadratic in fermions agree with previous calculations
[\BergshoeffFFOUR], while the quartic fermion terms have not previously been
given in the literature.
We want to stress the fact that the calculation is exact at this order.
It does not in any sense assume that ``$DF$ is small'', a kind of
assumption that is consistent in an abelian theory where it allows a
consistent truncation to slowly varying fields, but not in a non-abelian
theory, where commutators of covariant derivatives give field strengths. 
Derivative corrections to the abelian Born--Infeld action are discussed in
refs. [\AndreevTseytlinBIDerCorr,\WyllardDbraneCorr], 
but for the non-abelian case
there seems to be no previous results known at order $\a'^2$ that
incorporate all possible derivative terms.

\section\Supersymmetry{Linear and non-linear supersymmetry}Our 
formalism is manifestly $\N=1$ supersymmetric, so there is in principle
no need to check invariance under linear supersymmetry. However, 
in this section we will investigate the possibility of having 
$\a'$-corrected non-linear supersymmetries. 
The presence of such symmetries 
would indicate that the non-abelian 
theory could be viewed as embedded into a theory where all 
supersymmetries are linearly realised. The appearence of non-abelian 
parameters in this case would require
the introduction of non-abelian $\theta$ coordinates in the latter theory
which is a delicate enterprise (for a discussion of this and related
issues, see ref. [\BergshoeffKAPPA]). 
Here, however, the parameter of the non-linear
supersymmetry takes values only in the abelian part of the gauge group,
so the non-linear supersymmetry found here, although indicating the 
possibility of embedding the theory into another theory,
does not mean that this latter theory should be expressible 
in terms of non-abelian coordinates. 

In order to be able to discuss the relation
between the linear and non-linear supersymmetries, we first derive the
former.
Since linear supersymmetry
transformations are defined as translations along the anti-commuting
directions of superspace, see \eg\ ref. [\WessBagger], they read, using a 
parameter $\e^{\a}$ with a 
tangent space spinor index, 
$\d_\e(\phi)_{\th=0}=-(\e^\a\Dhat_\a\phi)_{\th=0}$.
Here we use the following conventions:
$\Dhat_\a$ is the ordinary flat superspace covariant derivative,
without any connection term, \ie, $\{\Dhat_\a,\Dhat_\b\}=-T_{\a\b}{}^c\*_c=
-2\G_{\a\b}^c\*_c$. The gauge-covariant derivatives are
$D_a=\*_a-A_a$, $D_\a=\Dhat_\a-A_\a$, and commute to
$[D_A,D_B\}=-T_{AB}{}^CD_C-F_{AB}$, where the connections and field 
strengths act in the appropriate representation (in this paper, all fields
carry the adjoint representation).
A gauge transformation with parameter $\Lambda$ is denoted $T[\Lambda]$. 

Of course, one may freely accompany $\d_\e$ by a gauge transformation.
For the multiplet at hand, it is convenient to add a gauge transformation
with parameter $\Lambda=\e^\a A_\a$, and define 
$Q_\e=\d_\e+T[\e^\a A_\a]$ in order to turn $\Dhat$ into a 
gauge-covariant derivative $D$. This leads to the following 
simple transformation
rules for the component fields (``$\th=0$'' suppressed):
$$
\eqalign{
Q_\e\cdot A_a&=\e^\a F_{a\a}=(\e\G_a\l)-7(\e\tJ_a)\komma\cr
Q_\e\cdot\l&=-\e^\a D_\a\l=\half(F_{ij}-\Fr{28}5K_{ij})\G^{ij}\e
	-\fr{24}(2K_{ijkl}+\Fr7{30}D^mJ_{mijkl})\G^{ijkl}\e
\punkt\cr
}\Eqn\ModifiedSS
$$
The supersymmetry algebra is ensured by the Bianchi identities. Consider
$[Q_\e,Q_{\e'}]A_a$. It is obtained as 
$$
\eqalign{
&Q_\e\cdot(\e'^\a F_{a\a})
-(\e\leftrightarrow\e')=-2\e^\a\e'^\b D_{(\a}F_{\vert a\vert\b)}\cr
&=-\e^\a\e'^\b(2\G^i_{\a\b}F_{ia}+D_aF_{\a\b})
=-2(\e\G^i\e')\*_iA_a+D_a\bigl(2(\e\G^i\e')A_i-\e^\a\e'^\b F_{\a\b}\bigr)
\komma\cr}
\eqn
$$
where the dimension 2 Bianchi identity is used in the second step,
so that the linear supersymmetries commute to a translation and a gauge
transformation for any $F_{\a\b}$,
$$
[Q_\e,Q_{\e'}]=-2(\e\G^i\e')\*_i
	+T[2(\e\G^i\e')A_i-\e^\a\e'^\b F_{\a\b}]\punkt\eqn
$$
On any tensor, the same algebra may of course be derived 
directly from the algebra
of $Q_\e=-\e^\a\Dhat_\a+T[\e^\a A_\a]=-\e^\a D_\a$.

For gauge groups containing a U(1) factor, the
undeformed action has a second, non-linearly realised supersymmetry:
$$
\left.
\eqalign{
\d_\h A_a&=0\cr
\d_\h\lambda&=\h\cr
}\right\}\quad
\Longrightarrow\quad[\d_\h,\d_{\h'}]=0\komma
\eqn
$$
with $\h$ taking values in the ${\frak u}(1)$ subalgebra.
We want to examine what happens to this symmetry when higher order
corrections are turned on.
Let us denote the above undeformed transformation $\d^{(0)}_\h$.
We then expect that $\d_\h$, if it remains unbroken by the
$\a'^2$-corrections, receives modifications $\a'^2\d^{(2)}_\h$.
Invariance means that $\d^{(0)}_\h\L'^{(2)}$ must be canceled against 
$$
\d^{(2)}_\h\L'^{(0)}
=\bigl(D_jG^{ji}+\half\{\x,\G^i\x\}\bigr)\d^{(2)}_\h B_i
	+\d^{(2)}_\h\x\Dslash\x\punkt
\Eqn\LzeroVar
$$
We therefore calculate $\d^{(0)}_\h\L'^{(2)}$, and find that it
is of the form (\LzeroVar) modulo total derivatives
This property is highly non-trivial, and relies on the precise numerical
coefficients in the action at order $\a'^2$ (it requires three linear
relations between the six coefficients of terms in $\L'$ containing $\x$).
The modified transformations\foot*{The terms in the transformations
containing $\ss D\hskip-5pt/\hskip1pt\x$ or $\ss D\hskip-5pt/\hskip1pt\l$ 
are only relevant for the
invariance of the action, not for the supersymmetry algebra, which only
closes on shell.  The linear supersymmetry
transformations  do not get corrected by such terms
 since they are derived from
the superspace formulation, and thus are on-shell transformations.}
are read off by comparing $\d^{(0)}_\h\L'^{(2)}$
to eq. (\LzeroVar):
$$
\eqalign{
\d_\h B_a&=-6\a'^2\Bigl[2G_{ai}(\h\G^i\x)-G^{ij}(\h\G_{aij}\x)\Bigr]\komma\cr
\d_\h\x&=\h-6\a'^2\Bigl[\fr2G^{ij}G_{ij}\h+\fr4G^{ij}G^{kl}\G_{ijkl}\h\cr
&-2(\h\G^i\x)D_i\x
	+\fr{15}(\h\G^{ijk}\x)\G_{ij}D_k\x\cr
&+\Fr{19}{25}(\h\G^i\x)\G_i\Dslash\x
	-\fr{90}(\h\G^{ijk}\x)\G_{ijk}\Dslash\x\Bigr]\komma\cr
}\Eqn\NLTransfI
$$
or, in terms of the original variables $A$ and $\l$:
$$ 
\eqalign{
\d_\h A_a&=-6\a'^2\Bigl[2F_{ai}(\h\G^i\l)-\fr6F^{ij}(\h\G_{aij}\l)\Bigr]
	\komma\cr
\d_\h\l&=\h-6\a'^2\Bigl[\fr{15}F^{ij}F_{ij}\h
	+\Fr2{15}F^{ij}F^{kl}\G_{ijkl}\h\cr
&-2(\h\G^i\l)D_i\l
	+\fr{15}(\h\G^{ijk}\l)\G_{ij}D_k\l\cr
&+\Fr{19}{25}(\h\G^i\l)\G_i\Dslash\l
	-\fr{90}(\h\G^{ijk}\l)\G_{ijk}\Dslash\l\Bigr]\punkt\cr
}\Eqn\NLTransfII
$$
The commutator of two non-linear supersymmetry transformations is 
read off directly from the transformations (\NLTransfI) or (\NLTransfII):
$$
\eqalign{
[\d_\h,\d_{\h'}]A_a&=-24\a'^2(\h\G^i\h')F_{ia}\komma\cr
[\d_\h,\d_{\h'}]\l&=-24\a'^2(\h\G^i\h')D_i\l\komma\cr
}\eqn
$$
modulo field equations.
Remember that the coeffients $M_{ABCD}$ are still present and suppressed.
Redefining the transformation parameter as 
$\r=2\sqrt{3\mu}\,\a'\h$, so that $\r$, like $\e$, has mass dimension
$-\half$, gives a
standard supersymmetry algebra:
$$
[S_\r,S_{\r'}]=-2(\r\G^i\r')\*_i
	+T[2(\r\G^i\r')A_i]
\komma\eqn
$$
provided that when we decompose the adjoint indices as $A=(0,A')$, 
0 denoting the U(1) factor, $M_{ABCD}$ satisfies 
$M_{0000}=\mu$, $M_{00A'B'}=\mu\d_{A'B'}$.
A straightforward but non-trivial
calculation also shows that the commutator between
a linear supersymmetry (from eq. (\ModifiedSS)) and a non-linear one 
yields a gauge transformation, 
$$
[Q_\e,S_\r]=T[-\fr{2\sqrt{3\mu}\a'}(\e\G_i\r)x^i
	+\Fr{\a'}{6\sqrt{3\mu}}(\e\G^{ijk}\r)(\l\G_{ijk}\l)]\komma\eqn
$$
modulo field equations
(we have satisfied ourselves with doing the calculation
acting on $A_a$). The shorthand notation of this equation is the same
as in the action: quadratic expressions are contracted with $\d_{AB}$,
so the first term is an U(1) gauge transformation, quartic ones with
$M_{ABCD}$, so the second one contains $M_{0ABC}$.
Unlike the ordinary (undeformed) 
super-Yang--Mills, where the non-linear supersymmetry
is abelian, the $\a'$-corrected theory may be seen as a theory with
partially broken $\N=(2,0)$ chiral supersymmetry
where $\l$ is the Goldstone fermion.
Starting from the $\N=(2,0)$ supersymmetry algebra and reintroducing
the $\a'$ rescaling of the $S$ generators, $\h\sim\a'^{-1}\r$, we can
understand the supersymmetry of the undeformed super-Yang--Mills theory
as a contraction as $\a'\rightarrow0$ of the $\N=(2,0)$ supersymmetry algebra.

\section\Structure{The structure of the order $\a'^2$ interactions}The 
tensor $M_{ABCD}$ is totally symmetric in the four adjoint indices.
Table 1 (last column) gives the number $P_4$ of linearly independent invariant
tensors of this kind for simple algebras. For all algebras, there
is always the tensor $\d_{(AB}\d_{CD)}$, and in the cases where the number
is 1, this is the only possible form of $M$. Algebras 
of ${\frak su}(N)$ type, $N\geq4$, have in addition
the tensor $d_{(AB}{}^Ed_{CD)E}$, and for $\frak sp$ 
and $\frak so$ algebras there is a
quartic invariant that can not be expressed in terms of a $d$-symbol with 
three indices (it can
be taken as 
$\Tr_f(T_{\mathstrut(A}T_{\mathstrut B}
T_{\mathstrut C}T_{\mathstrut D)})$). 
${\frak so}(8)$ has one more quartic 
invariant corresponding to the pfaffian.
Linear supersymmetry does not restrict $M$ further, so in the cases where 
$P_4>1$ the coefficients of the different possible invariants are 
unrelated.

If the algebra is not simple, more possibilities arise.
Of particular interest is the situation when 
${\frak g}={\frak g}'\oplus{\frak u}(1)$,
which is the case for multiple branes, 
${\frak g}={\frak u}(N)\simeq{\frak su}(N)\oplus{\frak u}(1)$.
Then, we decompose the adjoint index as $A=(0,A')$, 0 denoting
the ${\frak u}(1)$ part, and the possibilities are
$$
\eqalign{
M_{A'B'C'D'}&=a\,d_{(A'B'}{}^{E'}d_{C'D')E'}+b\,\d_{(A'B'}\d_{C'D')}\komma\cr
M_{0A'B'C'}&=c\,d_{A'B'C'}\komma\cr
M_{00A'B'}&=d\,\d_{A'B'}\komma\cr
M_{000A'}&=0\komma\cr
M_{0000}&=\mu\punkt\cr
}\Eqn\AllTheMs
$$
Any values of $a,b,c,d,\mu$ are consistent with linear supersymmetry.
If one in addition demands the second non-linearly realised supersymmetry,
one has to, as we saw in the previous section,
take $d=\mu\neq0$ in order for these
supersymmetries to commute to a translation. The remaining constants 
are unspecified. The ``symmetric trace'' prescription of Tseytlin
[\TseytlinSTR], used without referring to string theory, becomes identical
to our 
$M_{ABCD}\propto\Tr(T_{\mathstrut(A}T_{\mathstrut B}
T_{\mathstrut C}T_{\mathstrut D)})$. However, in string theory the trace
is by necessity in the fundamental representation which implies
that, for ${\frak su}(N)$, the generators satisfy 
$$
T_{A'}T_{B'}=\fr{N} \d_{A'B'}+\fr{2} ({d_{A'B'}}^{C'}+
{f_{A'B'}}^{C'})T_{C'}\komma\eqn
$$ 
and hence $M_{ABCD}$
contains a specific combination of all terms in eq. (\AllTheMs):
$a=1$, $b=\Fr4N$, $c=\Fr2{\sqrt N}$, $d=\mu=\Fr4N$.
 These restrictions are consistent with
the condition we obtained from non-linear supersymmetry namely $\mu=d$. 

\vskip3\parskip
\noindent
\vbox{\tabskip=0pt\offinterlineskip\def\ttt{\noalign{\hrule}}
\halign{#&\vrule#\tabskip=4pt&#\hfil&\vrule width1pt#
        &\hbox to 1cm{\hfil#\hfil}&\vrule#
        &\hbox to 1cm{\hfil#\hfil}&\hfil\vrule#
        \tabskip=0pt\cr\ttt
&height1pt&&&&&&\cr
&height11pt&Lie Algebra $\frak g$&&$P_3$&&$P_4$&\cr
&height4pt&&&&&&\cr\noalign{\hrule height1pt}
&height2pt&&&&&&\cr
&height11pt&$A_1$&&-&&1&\cr
&height11pt&$A_2$&&1&&1&\cr
&height11pt&$A_n$, $n\geq3$&&1&&2&\cr
&height4pt&&&&&&\cr\ttt
&height2pt&&&&&&\cr
&height11pt&$B_n$, $n\geq2$&&-&&2&\cr
&height4pt&&&&&&\cr\ttt
&height2pt&&&&&&\cr
&height11pt&$C_n$, $n\geq3$&&-&&2&\cr
&height4pt&&&&&&\cr\ttt
&height2pt&&&&&&\cr
&height11pt&$D_4$&&-&&3&\cr
&height11pt&$D_n$, $n\geq5$&&-&&2&\cr
&height4pt&&&&&&\cr\ttt
&height2pt&&&&&&\cr
&height11pt&exceptional&&-&&1&\cr
&height4pt&&&&&&\cr\ttt
}}

\noindent{\it Table 1. The number of singlets in the totally symmetric product
of 3 and 4 adjoints (could be enlarged to include higher Casimirs).}

\vfill\eject

\section\conclusions{Conclusions and comments}In this paper 
we have derived the $\a'^2$-corrections (linear supersymmetry does not allow 
any corrections at order $\a'$ [\CNTi]) 
to the supersymmetric Yang--Mills theory in ten dimensions.
The corrections obtained 
are the most general ones compatible with linear supersymmetry
which is implemented by the use of superspace. As shown in previous work
[\NilssonSYM,\CNTi] the superspace Bianchi identities 
can be solved in full generality
in terms of the component fields of a self-dual five-form superfield
$J_{abcde}$. By expressing this superfield in terms of the physical fields,
$F_{ab}$, $\l^{\a},$ and covariant derivatives, $D_a$, we derived in sections
\Interactions\ and \TheAction\ 
the action resulting from the $\a'^2$ terms in $J_{abcde}$ not 
removable by field redefinitions of the superfield $A_{\a}$. The action
includes quartic fermion terms not previously given in the literature as
well as a precise account of the freedom left after imposing (linear) 
supersymmetry.

This remaining freedom is given by 
$M^A{}_{BCD}$
which is symmetric in all four adjoint group 
indices and should be constructed from
the invariant group tensors $\d_{AB}$, $d_{ABC}$ and $f_{ABC}$. This is
elaborated upon in section \Structure. Each independent way of writing
$M^A{}_{BCD}$ corresponds to a new supersymmetric invariant that most likely 
will be given by an
infinite power series in $\a'$. This phenomenon will probably repeat
itself at higher orders producing leading terms in an infinite set of 
new invariants, but it should be stressed that it can not be excluded
that no additional freedom (\ie, new invariants) arises at higher orders. 
It is also important
to note that the non-linearities of the theory will in general 
cause the superinvariants
to mix at non-leading orders.  

However, the derivation of the action order by order in $\a'$ is a rather 
lengthy iterative process that contains some interesting aspects that could
change this picture. One such feature of the solution to the superspace Bianchi
identities obtained in [\CNTi] is that it requires the the tensor field
in the irreducible representation (00030) of $Spin(1,9)$ at first
order in the $\theta$ expansion of $J_{abcde}$ to vanish. 
This is the only non-trivial test of a non-vanishing $F_{\a\b}$ defining
the theory.
This starts to affect the analysis  at order $\a'^3$ where one has to make sure
that the terms that are allowed by group theory actually vanish. At 
order $\a'^4$, one has in addition to prove that the terms that do not vanish
are exactly the ones that cancel the $\a'^4$ terms found in the analysis 
of the (00030) condition in this paper (see comments in section {\old2}).
This analysis may turn out to provide rather severe restrictions at 
higher orders. Also, as we will discuss in a forthcoming publication 
[\CNTHigherOrder],
although there is a rather large number of possible terms contributing to
 $J_{abcde}$
at order $\a'^3$ most of them seem to be removable by field redefinitions
of $A_{\a}$ similar to the ones utilised in this paper at order $\a'^2$.
This amounts to identifying elements in the spinorial cohomology
discussed in refs. [\CNTi,\CNTCohomology] 
when expressed in terms of fields in the
vector multiplet.

Although the Born--Infeld action for a constant and abelian field strength
does not contain any terms of odd powers of $\a'$, such terms will most likely 
appear 
as soon as either of these restrictions is lifted. Derivative corrections to
the abelian theory have been discussed by several authors, see, \eg, refs.
[\AndreevTseytlinBIDerCorr,\CornalbaAbelBICorr,\WyllardDbraneCorr]  
showing that $\a'^3$ terms actually do not arise. In the non-abelian case,
the situation is unclear even for the $\a'^3$ order terms (an early superstring
computation of such terms can be found in ref. [\Kitazawa]). 
Applying our methods
to this case will hopefully clarify the situation.

Another intriguing feature of the corrections obtained here is that they
allow for a second non-linearly realised
supersymmetry provided, as explained in section 
\Supersymmetry, only a minor
restriction is imposed on $M^A{}_{BCD}$. This restriction is compatible with
the one obtained from string theory. It would be very interesting to 
investigate
how this non-linearly realised supersymmetry 
generalises to higher orders, and to see if it
can be understood as resulting from embedding the theory in a similar
 theory where all supersymmetries are
linearly realised along the lines of ref. [\BaggerGalperin]. 
For non-abelian theories such embeddings have not yet
been studied in any detail. However, some results that might be 
related to non-linearly realised symmetries with non-abelian valued parameters 
can be found in refs.
[\DouglasNCC,\MyersDIEL,\BergshoeffKAPPA]. 
What our work explains is how
a non-linear supersymmetry with an abelian parameter 
can be made to act on non-abelian fields
in a non-trivial manner extending 
the linear symmetry to an $\N=2$ supersymmetry.
It is clear that from the perspective of $\N=2$ supersymmetry, the
corrections at order $\a'^2$ considered in the present paper are
special---it is exactly this modification that changes the algebra
from a ``trivial'' abelian shift in the spinor to an ordinary supersymmetry
algebra, indicating that the theory may be embedded in a theory with linear
$\N=2$ supersymmetry with the ${\frak u}(1)$ part of $\l$ as the Goldstone
fermion. We find it very striking that the requirement of non-linear
supersymmetry contains practically no information on the structure
of the interactions not already implied by linear supersymmetry.
It will be interesting to examine whether this statement continues to
hold at higher orders.

\appendix{Spinors and Fierz identities}We use real chiral (Majorana-Weyl)
spinors throughout the paper. The two chiralities are distinguished by
subscript for the (00010) representation (\eg\ $D_\a$) and superscript
for the (00001) (\eg\ $\l^\a$). The $\G$-matrices are thus not Dirac
matrices but ``Pauli matrices''; we use the same notation $\G$ for
$\G^a_{\a\b}$ and $\G^{a\,\a\b}$.

In deriving the final form of the equations of motion one needs to rearrange
a number of trilinear $\l$ terms. For this purpose the following Fierz 
identities are quite useful.
The general identity
$$
(\l A\{\l),B\l\}
	=\fr{16}\{\l,\G^i\l\}B\G_i A^t\l
		+\fr{32\cdot5!}\{\l,\G^{ijklm}\l\}B\G_{ijklm}A^t\l
\Eqn\GeneralSymFierz
$$
leads to the various Fierz identities
$$
\eqalign{
(\l\G_{[a}\{\l),\G_{b]}\l\}
	&=\fr{16}\{\l,\G^i\l\}\G_{abi}\l
		+\fr{96}\{\l,\G_{abijk}\l\}\G^{ijk}\l\komma\cr
(\l\G^i\{\l),\G_{abi}\l\}
	&=\{\l,\G_{[a}\l\}\G_{b]}\l-\Fr38\{\l,\G^i\l\}\G_{abi}\l
		+\fr{48}\{\l,\G_{abijk}\l\}\G^{ijk}\l\komma\cr
(\l\G_{abi}\{\l),\G^i\l\}
	&=\{\l,\G_{[a}\l\}\G_{b]}\l+\Fr38\{\l,\G^i\l\}\G_{abi}\l
		-\fr{48}\{\l,\G_{abijk}\l\}\G^{ijk}\l\komma\cr
(\l\G_{ij[a}\{\l),\G^{ij}{}_{b]}\l\}
	&=\Fr74\{\l,\G^i\l\}\G_{abi}\l
		-\fr{24}\{\l,\G_{abijk}\l\}\G^{ijk}\l\komma\cr
(\l\G^{ijk}\{\l),\G_{abijk}\l\}
	&=42\{\l,\G_{[a}\l\}\G_{b]}\l-\Fr{21}4\{\l,\G^i\l\}\G_{abi}\l
		-\Fr38\{\l,\G_{abijk}\l\}\G^{ijk}\l\komma\cr
(\l\G_{abijk}\{\l),\G^{ijk}\l\}
	&=42\{\l,\G_{[a}\l\}\G_{b]}\l+\Fr{21}4\{\l,\G^i\l\}\G_{abi}\l
		+\Fr38\{\l,\G_{abijk}\l\}\G^{ijk}\l\punkt\cr
}\eqn
$$
Eq. (\GeneralSymFierz)
is most easily derived by
recalling that $(\l A\{\l),B\l\}$ means 
$(\l^AA\l^B)(B\l^C)_{\a}-(\l^AA\l^C)(B\l^B)_{\a}$ (the $B$ and $C$ indices
are contracted by a structure constant, but it is only the symmetry that is
relevant here),
and then use the following expansion of the product of 
two spinors anti-symmetrised in the adjoint indices and hence 
symmetrised in the spinor indices:
$$
\l^{[B}_{\a\mathstrut}\l^{C]}_{\b\mathstrut}=\fr{16}(\G^a)_{\a\b}(\l^B\G_a\l^C)
+\fr{32\cdot5!} (\G^{abcde})_{\a\b}(\l^B\G_{abcde}\l^C)\komma\eqn
$$
where the coefficients follow directly from the  trace formul\ae\ for
chirally projected 16 by 16 $\G$-matrices:
$$
(\G^a)^{\a\b}(\G^b)_{\a\b}=16\eta^{ab}\komma\eqn
$$
and 
$$
\eqalign{
(\G^{a_1...a_5})^{\a\b}(\G_{b_1...b_5})_{\a\b}&=
tr(\fr{2}(\hat{1}+\hat{\G}^{11})\hat{\G}^{a_1...a_5}\hat{\G}^{b_1...b_5})\cr
&=16\cdot5!\delta^{a_1...a_5}_{b_1...b_5}+
16{\epsilon^{a_1...a_5}}_{b_1...b_5}\komma\cr
(\G^{a_1...a_5})_{\a\b}(\G_{b_1...b_5})^{\a\b}&=
tr(\fr{2}(\hat{1}-\hat{\G}^{11})\hat{\G}^{a_1...a_5}\hat{\G}^{b_1...b_5})\cr
&=16\cdot5!\delta^{a_1...a_5}_{b_1...b_5}-
16{\epsilon^{a_1...a_5}}_{b_1...b_5}\komma\cr
}\Eqn\chiraltraces
$$
where $tr$ refers to traces over non-chiral 32 by 32 $\hat{\G}$ matrices.

Other useful Fierzes, involving expressions like
$(\l\G^iD^j\l)\G_iD_j\l$ which with an explicit $M$ becomes
${M^A}_{BCD}(\l^B\G^iD^j\l^C)\G_iD_j\l^D$, are:
$$
\eqalign{
(\l\G^iD^j\l)\G_iD_j\l&=\fr{24}(D^l\l\G_{ijk}D_l\l)\G^{ijk}\l\komma\cr
(\l\G^iD^j\l)\G_jD_i\l&=\fr{48}(D^l\l\G_{ijk}D_l\l)\G^{ijk}\l
		+\fr4(D_i\l\G_jD_k\l)\G^{ijk}\l\komma\cr
(\l\G^{ijk}D^l\l)\G_{ijk}D_l\l&=-\fr2(D^l\l\G_{ijk}D_l\l)\G^{ijk}\l\komma\cr
(\l\G^{ijk}D^l\l)\G_{ijl}D_k\l&=-\fr{24}(D^l\l\G_{ijk}D_l\l)\G^{ijk}\l
		+\Fr52(D_i\l\G_jD_k\l)\G^{ijk}\l\komma\cr
}\Eqn\ldldlfierz
$$
where we have used also the lowest order field equation $\Dslash\l=0$.

\acknowledgements
This work is partly  
supported by EU contract HPRN-CT-2000-00122 and by the Swedish
Natural Science Research Council.

\refout

\end